\documentclass[prb, amsmath, reprint, amssymb, aps, superscriptaddress]{revtex4-1}
\usepackage{graphicx}
\usepackage{amsmath, braket, amsfonts}

\usepackage{multirow}
\usepackage{mathrsfs,amsmath} 
\usepackage{fixmath}
\usepackage{graphicx}
\usepackage[colorlinks,linkcolor=blue,anchorcolor=red,citecolor=green]{hyperref}
\usepackage{placeins}

\renewcommand\vec{\mathbf}

\begin{document}


\title{Quantum paracrystalline shear modes of the electron liquid}

\author{Jun Yong Khoo}
\affiliation{Max-Planck Institute for the Physics of Complex Systems, D-01187 Dresden, Germany}

\author{Po-Yao Chang}
\affiliation{Department of Physics, National Tsing Hua University, Hsinchu 30013, Taiwan}
\affiliation{Max-Planck Institute for the Physics of Complex Systems, D-01187 Dresden, Germany}

\author{Falko Pientka}
\affiliation{
Institute of Theoretical Physics, Goethe University, 60438 Frankfurt a.M., Germany
}
\affiliation{Max-Planck Institute for the Physics of Complex Systems, D-01187 Dresden, Germany}

\author{Inti Sodemann}
\affiliation{Max-Planck Institute for the Physics of Complex Systems, D-01187 Dresden, Germany}


\date{\today}

\begin{abstract}
Unlike classical fluids, a quantum Fermi liquid can support a long-lived and propagating shear sound wave at arbitrarily small wave vectors and frequencies, reminiscent of the transverse sound in crystals, despite lacking any form of long-range crystalline order. This mode is expected to be present in moderately interacting metals where the quasiparticle mass is renormalized to be more than twice the bare mass in two dimensions (2D), but it has remained undetected because it is hard to excite since it does not involve charge density fluctuations, in contrast to the conventional plasma mode. In this work we propose a strategy to excite and detect this unconventional mode in clean metallic channels. We show that the shear sound is responsible for the appearance of sharp dips in the ac conductance of narrow channels at resonant frequencies matching its dispersion. The liquid resonates while minimizing its dissipation in an analogous fashion to a sliding crystal. 
Ultra-clean 2D materials that can be tuned towards the Wigner crystallization transition such as silicon metal-oxide-semiconductor field-effect transistors, MgZnO/ZnO, p-GaAs, and AlAs quantum wells are promising platforms to experimentally discover the shear sound.
\end{abstract}


\maketitle

\section{Introduction}

Ordinary classical fluids only display one kind of sound waves that correspond to longitudinal compressional oscillations of the fluid~\cite{LLfluids}.
On the other hand classical solids display transverse waves as well, which originate from their finite restoring force to shear deformations~\cite{LLelasticity}.
Quantum Fermi fluids can dramatically differ from this paradigm by displaying long-lived and propagating collective shear sound waves at arbitrarily small frequency and wave vector while lacking any form of static crystalline order~\cite{Pines, Conti, Shear, Chubukov, Alekseev2019b}.

To this date there is no report of the observation of these shear sound waves of electrons in metals, and a pioneering attempt to detect them in $^3$He~\cite{Roach}
remained inconclusive~\cite{Flowers}.
However, the appearance of these modes requires only a moderate interaction strength, in the sense that they are expected to become sharp when the quasiparticle mass becomes approximately twice and three times the transport mass in two- and three-dimensions respectively~\cite{Shear}. Therefore it is possible that these elusive collective modes are actually present in a variety of electron liquids but they have remained undetected so far because their transverse nature makes them unresponsive to charge-sensitive probes.

\begin{figure*}[t]
\includegraphics[scale=1.0]{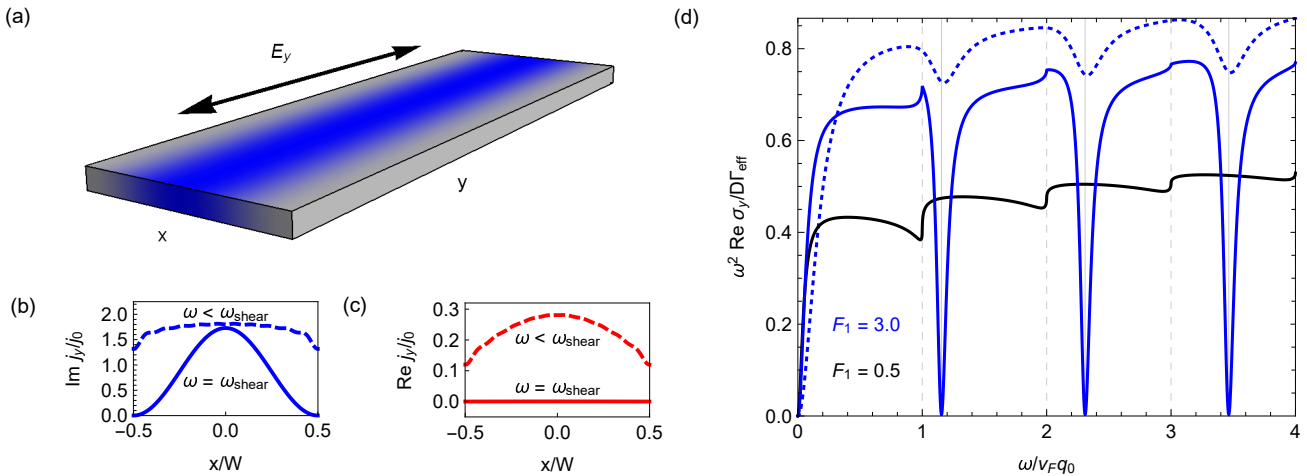}
\caption{
(a) Experimental setup to detect shear sound. The blue region illustrates the out-of-phase (imaginary) current profile in the channel. (b) Out-of-phase (imaginary) and (c) in-phase (real) current profiles for driving frequency on and off resonance with the shear sound frequency $\omega _{\rm shear}$. 
(d) Real part of the transverse conductivity in units of $D \Gamma _{\rm eff} / \omega ^2$ when the shear sound is present (solid blue line) and absent (solid black line) in the limit of boundary dominated scattering [boundary scattering parameter $b = 0.1 (2 \pi v_{\rm F})$], where $D = ne^2/m$ is the Drude weight and $\Gamma _{\rm eff}$ is an effective scattering rate~\cite{supplementary}. For finite bulk scattering ($\Gamma _1 = 0.1 v_{\rm F}q_0$), the resonant zeros at the shear sound harmonics (solid blue line) evolve into smooth dips (dashed blue line). 
}\label{Fig.mainfig}
\end{figure*}

In this paper, we demonstrate that shear modes leave clear fingerprints in the conductivity of clean metallic channels. 
Our idealized setup is depicted in Fig.~\ref{Fig.mainfig}(a), where a uniform ac electric field generates an alternating current along the $y$ direction. In a clean channel, the current can only be damped at the boundary. 
This is illustrated by the current profile shown in Figs.~\ref{Fig.mainfig}(b) and \ref{Fig.mainfig}(c), which is suppressed at the boundaries due to friction. The current magnitude varies in a direction \textit{transverse} to the electron flow signaling the excitation of shear modes. 

The central result of our work is summarized in Fig.~\ref{Fig.mainfig}(d), which shows the conductance of the strip as a function of frequency. When scattering due to impurities or electron-electron collisions is weak, the conductance exhibits sharp dips at frequencies $\omega=n \omega_0$, where $\omega_0$ is the shear sound frequency at momentum $2\pi/W$ determined by the width $W$ of the channel. 
In fact, when friction only occurs at the boundary (blue curve), the conductivity vanishes on resonance and the liquid responds in a dissipationless fashion. 
As we will show, this is a characteristic transverse response of a sliding crystal which is only subjected to friction at the boundaries.  Therefore these resonances reveal a type of crystallinity that appears in Fermi liquids when probed dynamically. Such remarkable collective behavior could be observed in ultra-clean samples such as those recently employed to observe the hydrodynamic electronic flow~\cite{electronhydro1,electronhydro2,electronhydro3,Alekseev2019} but in the low-temperature quantum regime where the classical hydrodynamic description breaks down. 
A related behavior in the form of oscillations of the absorption power as a function of magnetic field was predicted in Ref.~\onlinecite{Alekseev2019} (see Fig.~2 of this reference). We note, however, that in the regime of long wavelengths in a magnetic field there is no well-defined separation into transverse and longitudinal modes leading to a crucially distinct regime of collective modes from the one studied here.

The conductivity dips shown in Fig.~\ref{Fig.mainfig}(d) are unique signatures of the shear sound that would be absent in weakly interacting metals where this mode does not exist (black curve). Likewise, the dips are washed out once scattering in the bulk becomes comparable to the boundary friction (dashed curve). This is a consequence of a reduced shear force when the force difference between the interior and the boundary is small as we will describe in detail.

Our paper is organized as follows. Section~\ref{Sec.II} generalizes the discussion of Ref.~\onlinecite{Shear} to describe the behavior of shear modes in the presence of impurity and electron-electron collisions in an ideal infinite two-dimensional (2D) system without boundaries. Section~\ref{Sec.III} is devoted to a conceptual discussion reviewing some of the key similarities and differences between the quantum Landau Fermi liquid (LFL), crystalline solids, ordinary classical fluids, and viscoelastic classical fluids, also for ideal infinite-size 2D systems. In Sec.~\ref{Sec.IV} we develop a theory to describe the hydrodynamics of the LFL in a strip geometry and derive the exact analytic solution which predicts the appearance of shear resonances in experiments. In Sec.~\ref{Sec.V} we show that these resonances are analogous to those arising from an ideal crystal sliding in a channel by studying a toy model. We summarize our results and discuss potential material candidates to observe these shear sound modes in Sec.~\ref{Sec.VI}.

\section{Diffusive and propagating shear modes}\label{Sec.II}

At low temperatures metals enter the quantum Landau Fermi liquid (LFL) regime. A Fermi liquid can be thought of as having an infinite number of slow degrees of freedom that describe the relaxation of the shape of the Fermi surface. Unlike superfluids or ordinary classical liquids, the low-energy excitations of LFLs cannot be captured completely by a description in terms of a finite number of dynamical fields such as density and current. We will focus on 2D systems but many of our conclusions carry over to the three-dimensional (3D) case.

We begin by stating a central finding of our study: even in the presence of collisions, 2D Fermi liquids display a sharp propagating transverse sound mode with speed $v_{\rm s}=v_{\rm F} (1+F_1)/2\sqrt{F_1}$, for Landau parameter $F_1>1$, and for wave vectors $q \gtrsim q_*$, with $q_* = \max \left\lbrace \Gamma_1/v_{\rm s}, \Gamma_2/v_{\rm F} \sqrt{F_1} \right\rbrace$, where $v_{\rm F}$ is the Fermi velocity and $\Gamma_1,\Gamma_2$ are the momentum-relaxing and -preserving collision rates, respectively. We will now derive these results within the Landau theory of Fermi liquids.

In LFL theory the shape of the Fermi surface becomes a dynamical object and small deviations of the radius $p_{\rm F}(r,\theta)$ from the equilibrium shape obey the linearized Landau kinetic equation (LKE)~\cite{Pines}:
\begin{align}
\partial _t p_{\rm F} (\vec{r}, \theta) &+ \vec{v}_p \cdot \vec{\partial}_{\vec{r}}\Bigl[ p_{\rm F} (\vec{r}, \theta)+ \int \frac{d \theta '}{2\pi} f(\theta - \theta ') p_{\rm F} (\vec{r}, \theta ') \Bigr]\nonumber \\
&= - e \vec{E} \cdot \vec{v}_p + I [p_{\rm F}]. \label{Eq.LKE}
\end{align}
Here, $\vec{v}_p = v_{\rm F}\hat{p}$ is the velocity normal to the Fermi surface at angle $\theta$, $f(\theta-\theta')$ is the Landau function including short-range and Coulomb interactions, $\vec{E}$ is the applied electric field, and $I$ are collision terms. There are two kinds of collisions terms: those which relax momentum, such as electron-impurity collisions, and those that preserve momentum, originating from electron-electron collisions, which can be modeled as~\cite{LevGregNat,LevGregPNAS,Alekseev2018}:
\begin{eqnarray}
I [p_{\rm F}] &=& - \Gamma _1 (p_{\rm F} - P_0[p_{\rm F}]) \nonumber \\
&&- \Gamma _2 (p_{\rm F} - P_0[p_{\rm F}] - P_1[p_{\rm F}] - P_{-1} [p_{\rm F}]).
\end{eqnarray}
Here, $P_m[p_{\rm F}]$ projects the Fermi radius onto the $m$th harmonic $e^{i m \theta}$.
There are two types of solutions to the LKE: incoherent and collective modes. The incoherent modes are sharply \textit{localized} angular deformations of the Fermi surface~\cite{Pines, Shear} that form the particle-hole continuum with a dispersion of the form:
\begin{equation}
\omega _{\rm p-h}= v_{\rm F} q  \cos \theta+i(\Gamma_1+\Gamma_2).
\end{equation}
Collective modes, however, are angularly \textit{delocalized} deformations of the Fermi surface~\cite{Pines, Shear}. When the system has a microscopic mirror symmetry and the wave vectors of the modes lie along the mirror invariant line, the modes can be separated into odd (transverse) and even (longitudinal) under the mirror operation~\cite{Pines, Shear}. The well-known plasma mode of metals is a longitudinal mode, whereas, the shear sound is a transverse mode.

To illustrate the features of the shear sound, we consider a simplified model in which all the $n > 1$ angular moments of the Landau interaction function vanish, $F_{n > 1} = \int (d \theta/2\pi) f(\theta ) \cos (n \theta) = 0$.  The $F_1$ parameter controls the ratio of the quasiparticle mass ($m^*$) to the Drude mass ($m$) of a Fermi liquid, $m^*=(1+F_1) m$. The Drude mass would equal the non-interacting mass ($m_0$) in Galilean invariant systems~\cite{Randeria, Baeriswyl, Varma, Okabe}.

Our key results are expected to remain valid in the presence of other Landau parameters whenever the shear sound mode remains the only sharp collective mode in the transverse sector~\cite{Shear}. For this model, a LFL with $F_1>1$ would feature a propagating shear sound mode with dispersion:
\begin{equation}\label{Eq.shearcomplexdispphys}
\omega _{\rm s} = i \left( \Gamma _1 + v_{\rm s} q_2\right) + v_{\rm s}\sqrt{q^2 - q_2^2}, \quad q_2 = \frac{\Gamma _2}{v_{\rm F} \sqrt{F_1}}.
\end{equation}
This mode exists for $q>q_2$, whereas for $q<q_2$ one encounters diffusive collective modes as depicted in Fig.~\ref{Fig.schematic}(a) and detailed in the Supplemental Material~\cite{supplementary}.
Therefore, the shear sound is expected to become a sharp collective mode in moderately interacting Fermi liquids ($F_1>1$) for $q > q_*$, with
\begin{equation}
q_* \approx \max \left\lbrace \frac{\Gamma_1}{v_{\rm s}}, q_2 \right\rbrace.
\end{equation}
In the $q_2 \ll q \ll p_{\rm F}$ limit, the shear sound velocity asymptotes to its undamped value $v_{\rm s}$~\cite{Shear}. 
On the other hand, for a weakly interacting LFL with $|F_1| < 1$, only a single, purely decaying collective mode exists as depicted in Fig.~\ref{Fig.schematic}(b), with dispersion:
\begin{eqnarray}\label{Eq.weakdiff}
\omega_{\rm diff} &=& i \left( \Gamma _1 +v_{\rm s} q_2- v_{\rm s} \sqrt{q_2^2 - q^2}\right) , \\
&\simeq & i \left(\Gamma _1 + \frac{v_{\rm F}}{2Q} q^2 + \mathcal{O}(q^4) \right), \quad Q = \frac{1}{v_{\rm F}}\frac{2\Gamma _2}{1+F_1}.
\end{eqnarray}
This decaying mode exists for $0 \leq q \leq Q$, where its relaxation rate increases with $q$ from $\Gamma _1$ at $q \rightarrow 0$ to $\Gamma_1+\Gamma_2$ at $q = Q$, 
as shown in Fig.~\ref{Fig.schematic}(b). We have found that $\Gamma_1+\Gamma_2$ is, within our model, the momentum-independent value of the decay rate of all the modes that make up the particle-hole continuum. Therefore, in the presence of collisions the particle-hole continuum is displaced as a whole to lie in a plane of constant imaginary part, and is depicted by the green region in Figs.~\ref{Fig.schematic}(a) and \ref{Fig.schematic}(b). Notice that this transverse mode becomes strictly diffusive only in the limit of vanishing momentum-relaxing collisions $\Gamma_1 \rightarrow 0$, and exists only for a non-vanishing rate of momentum preserving collisions $\Gamma_2 > 0$. Therefore, at such small wave vectors the weakly interacting Fermi liquid ($|F_1|<1$) behaves like a classical fluid, as we will describe in more detail in the next section, where the slow diffusive relaxation of transverse currents is a consequence of the local conservation of momentum~\cite{LLfluids}.
When $F_1 < -1$, one finds instead exponentially growing modes associated with a Pomeranchuk instability~\cite{Chubukov, Chubukovmirage, supplementary}.

\begin{figure}[h]
\includegraphics[scale=1.0]{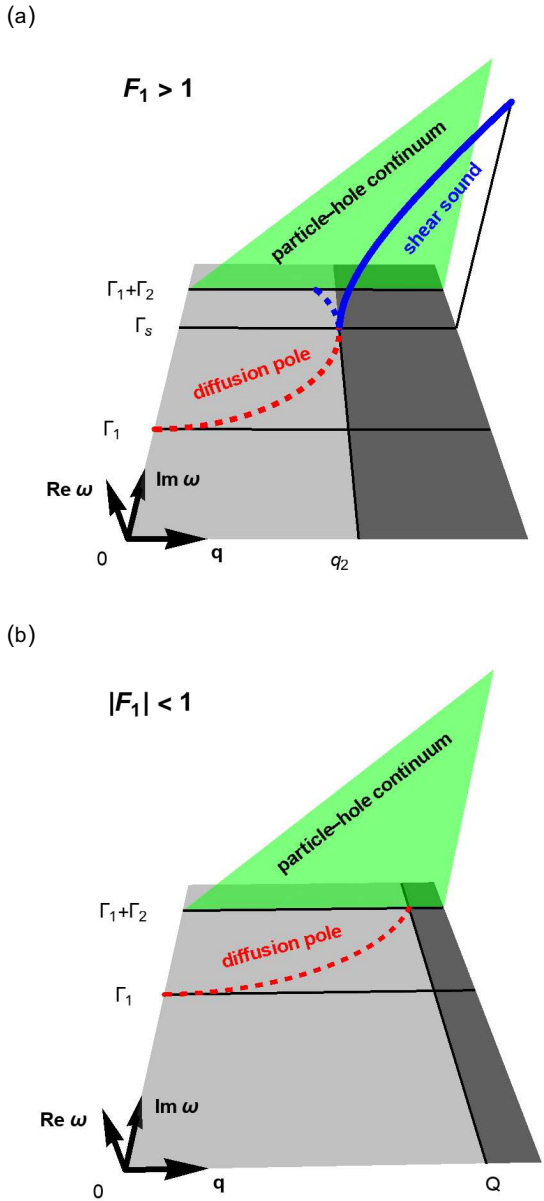}
\caption{
(a) The dispersive shear sound (blue solid curve) exists only for moderately interacting Fermi liquids ($F_1 > 1$) and relaxes at a lower rate than that of the incoherent particle-hole excitations (green wedge), $\Gamma _s = \Gamma _1 + v_{\rm s} q_2 < \Gamma _1 + \Gamma _2$. 
(b) The dispersive shear sound is absent when interactions are too weak ($F_1 < 1$).
Red and blue dashed curves indicate the dispersion of decaying collective shear modes.}\label{Fig.schematic}
\end{figure}

\section{Transverse modes in Fermi liquids, classical viscoelastic fluids and crystalline solids}\label{Sec.III}

In this section, we would like to discuss the relations between the transverse current responses of the quantum Fermi liquid, classical fluids, and crystalline solids.  We review some remarkable similarities but also sharp differences between these systems and the quantum LFLs at small $(q, \omega)$. This serves as a reminder that analogies between quantum LFLs and classical states of matter must be employed cautiously even in the limit of small $(q, \omega)$, and that these systems ultimately belong to different universality classes. For conceptual clarity we restrict the discussion in this section to translationally invariant fluids by taking the momentum-relaxing rate to be $\Gamma _1 = 0$ from the outset.

We would like to begin by making precise what we mean by ``quantum'' in ``quantum LFL''. When we refer to a ``quantum LFL'' we are emphasizing that this 
is a state of matter which is strictly speaking only well defined at $T=0$, although its consequences permeate to finite temperatures, analogous to the terminology employed in quantum critical phenomena. Therefore, the long-wavelength response of the ``quantum LFL'' is defined by taking first the limit $T\rightarrow 0$, and then afterwards taking the limits of small $(q, \omega)$. This order is crucial as the two limits do not commute. In fact, in the opposite case when $(q, \omega) \rightarrow 0$ while keeping temperature fixed, the response of the LFL is identical to that of an ordinary classical fluid, as well shall see below. In the language of critical phenomena, temperature can be viewed as a relevant perturbation that transforms the universal properties of the liquid at sufficiently long wavelengths. In our formalism, temperature enters through the momentum-preserving quasiparticle collision rate, which scales with temperature as $\Gamma_2\sim\left(k_B T\right)^2/E_{\rm F}$
up to logarithmic corrections~\cite{Hodges1971,Chaplik1971,Bloom1975,eecolrate,Fukuyama1983,Zheng1996,Jungwirth1996,
Menashe1996,Reizer1997,Narozhny2002,novikov2006viscosity}.

In Sec.~\ref{Sec.II}, we have seen that at $T=0$ the shear sound is indeed a sharp linearly dispersing mode at small $(q, \omega)$, reminiscent of solids also featuring a propagating shear sound at arbitrarily small $(q, \omega)$ but unlike classical fluids (including viscoelastic fluids), which display shear diffusion at small $(q, \omega)$. At finite temperature and sufficiently small $(q, \omega)$, LFLs also exhibit a shear diffusion mode. To make these similarities and distinctions more concrete, we will review the limiting behavior of the transverse conductivity for these various states of matter in the remainder of this section.

We begin by considering the case of an ordinary classical liquid with the same symmetries as the quantum Fermi liquid we are interested in: homogeneity, isotropy, time reversal, etc. Such liquids can be described at long wavelengths by the Navier-Stokes equation~\cite{LLfluids}, which upon linearization yields the transverse conductivity (that measures the current density in response to an external transverse force) 
\begin{equation}\label{Eq.transcondCLfull}
\sigma _\perp ^{\rm CL} (q, \omega) = \frac{n e^2}{m} \frac{1}{i \omega + \frac{\eta}{mn}q^2},
\end{equation}
where $\eta$ is the shear viscosity of the liquid. As we see, there is a diffusive pole for transverse currents, with diffusion constant $D = \eta / m n$.
%

Now, let us consider an ordinary crystalline solid which at long distances has also the same symmetries of interest. We take the solid to be described by an effective elasticity theory, from which the conductivity can be easily derived by adding an external force to the elasticity equations of motions~\cite{LLelasticity}. In particular, the transverse conductivity,
\begin{equation}\label{Eq.transcondCS}
\sigma _\perp ^{\rm CS} (q, \omega) = \frac{n e^2}{m} \frac{1}{i \left(\omega - c_t^2 \frac{q^2}{\omega}\right)},
\end{equation}
features a real and linearly dispersing pole at $\omega = c_t q$, signaling the presence of a propagating transverse sound mode in the solid. The transverse sound velocity $c_t$ can be related to the shear modulus $\mu$ of the solid as $c_t^2 = \mu / m n$. 

Let us now consider the transverse response of the quantum LFL. The full expression of the transverse conductivity of the 
bulk Fermi liquid will be presented in Eq.~\eqref{Eq.condbk} and here we present its zero temperature and clean limit ($\Gamma _{1,2} = 0$) at small $\omega$ and $q$ but with an arbitrary ratio of $s = \omega / v_{\rm F} q$:
%
%
%
\begin{equation}\label{Eq.transcond0}
\sigma _\perp ^{\rm LFL} (q, \omega, T = 0) = \frac{n e^2}{m} \frac{2}{i v_{\rm F}q} \frac{s - \sqrt{s^2 - 1}}{1 - F_1 (s - \sqrt{s^2-1})^2}.
\end{equation}
where the frequencies lie in the lower half plane, $s\to s-i0$. The response has nonanalyticities at the onset of the particle-hole continuum of incoherent excitations at $\omega = v_{\rm F} q$. This threshold is ultimately a consequence of the existence of an underlying sharp Fermi surface. It is easy to verify that when $F_1 >1$ the denominator of the transverse conductivity of the quantum Fermi liquid has a zero at the ideal $T=0$ dispersion of the shear sound mode~\cite{Shear} corresponding to the $\Gamma _{1,2} \rightarrow 0$ limit of Eq.~\eqref{Eq.shearcomplexdispphys}:
\begin{equation}\label{Eq.shear0}
\omega _{\rm s} (\Gamma _{1,2} = 0) = v_{\rm s} q, \quad v_s =  \frac{1 + F_1}{2\sqrt{F_1}} v_{\rm F}.
\end{equation}
Notice that 
the condition $F_1 >1$, is precisely 
that which needs to be satisfied so that the speed of the shear sound $v_{\rm s}$ is larger than $v_{\rm F}$, which is a self-consistent requirement if it is to be a well-defined propagating mode outside of the particle-hole continuum.

\begin{table}[t]
\begin{tabular}[b]{ccccccc}
\hline\\[-1.em]
\hline\\[-1.em]
& &Liquid &\quad \quad \quad &Solid &\quad \quad \quad &LFL ($T=0$)\\ [.5em]
\hline\\[-1.em]
$\displaystyle \lim_{\omega \rightarrow 0} \sigma _\perp (q, \omega) $ & &  $\displaystyle \frac{n^2 e^2}{m \eta q^2}$ & &$0$ & & $\displaystyle \frac{e^2}{h} (2S + 1) \frac{p_{\rm F}}{q}$\\ [1.em]
\hline\\[-1.em]
\hline\\[-1.em]
\end{tabular}
\caption{Quasistatic limit of the transverse conductivity in classical liquids, crystalline solids, and zero temperature LFLs. The factor $(2S + 1)$ is the spin degeneracy factor of the Fermi fluid (2 for usual spin-$\frac{1}{2}$ fermions).}\label{table}
\end{table}

%

To compare the transverse response of these different systems we first consider the ``optical'' regime $\omega \gg v_{\rm F} q$, where a quantum Fermi liquid resembles a solid as emphasized in the seminal work of Conti and Vignale~\cite{Conti}. Indeed, the transverse response of the quantum Fermi liquid in this regime
\begin{equation}\label{Eq.transcondLFL}
\sigma _\perp ^{\rm LFL} (\omega \gg v_{\rm F} q, T = 0) \approx \frac{n e^2}{m} \frac{1}{i\left( \omega - \frac{1 + F_1}{4} v_{\rm F}^2 \frac{q^2}{\omega}\right)}
\end{equation}
is identical to that of the crystalline solid in Eq.~\eqref{Eq.transcondCS}. 
When $F_1\gg 1$ the above form has a pole inside the optical regime at $\omega = \sqrt{1 + F_1}v_{\rm F} q/2$, which corresponds to the ideal shear sound dispersion from Eq.~\eqref{Eq.shear0} in that limit, and which was first obtained in Ref. ~\citenum{Conti}. 

Notice that the expansion in Eq.~\eqref{Eq.transcondLFL}, when extrapolated without caution, appears to indicate that the Fermi liquid always has a shear sound mode. 
However, as we have seen, the shear sound mode only appears as a separate mode for $F_1>1$.
For intermediate values of $F_1$ the analogy between quantum LFLs and crystalline solids fails because particle-hole excitations cannot be ignored. In particular, the transition at $F_1 = 1$, where the shear sound merges with the particle-hole continuum, cannot be captured by a classical fluid or elasticity theory.

The difference between classical and quantum regime is most striking in the quasi-static limit $\omega\ll v_F q$, where the quantum response is dominated by the particle-hole continuum. The transverse response in this regime for the three different cases is listed in Table~\ref{table}. While $\sigma _\perp$ has a $1/q^2$ dependence in a liquid, a solid cannot flow when subjected to static perturbations and exhibits a vanishing transverse conductivity at $\omega=0$. In contrast, the quantum Fermi liquid has a remarkable universal form $\propto 1/q$ in the quasi-static limit. The limit is finite in contrast to the solid, because the Fermi liquid still flows, but it is distinct from that of a classical fluid. This limiting response of the quantum LFL is universal in the sense that it is not renormalized by interactions and only depends on the geometry of the Fermi surface~\cite{Pines}. Notice also the appearance of Planck's constant in the denominator, a reminder of the quantum nature of the response in this limit. We will elaborate on the physics and measurement of this limit in a forthcoming publication and demonstrate that another quantum fluid, the spinon Fermi surface, which also features a sharp Fermi surface despite not being a LFL, has the same behavior in this limit.

While the quantum Fermi liquid at strictly $T=0$ is clearly distinct from solids and classical fluids, finite temperatures smear out the sharpness of the Fermi surface on a scale $k_BT/v_F$, destroying the ``quantumness'' of the fluid at sufficiently small $q$. In the following, we elucidate how the classical behavior is recovered in LFL theory once the limit of small $q$ is taken at finite temperature.

A useful point of comparison for LFLs at finite temperatures are classical viscoelastic fluids, which can also display long-lived shear modes~\cite{LLelasticity, Dyre2006, Trachenko2015, Lucas2019, Trachenko2020}. Specifically, we focus on the Frenkel model often employed in the description of classical viscoelastic fluids~\cite{Trachenko2015, Lucas2019, Trachenko2020}. Following Refs.~\citenum{Trachenko2015} and~\citenum{Trachenko2020}, we add to the Navier-Stokes-Frenkel equation an external force per unit area $\vec{f} = n e \vec{E}$ to obtain the equation of motion
\begin{equation}
\eta  \vec{\partial}_{\vec{r}}^2 \vec{v} = (1 + \tau d_t ) (n m d_t \vec{v} + \vec{\partial}_{\vec{r}} p - \vec{f}),
\end{equation}
where $d_t = \partial _t + \vec{v}\cdot \vec{\partial}_{\vec{r}}$. Upon linearizing this equation one finds that the transverse current $j_{\perp} =n e v_{\perp}$ has an associated transverse conductivity 
\begin{equation}\label{Eq.transcondFr}
\sigma _\perp ^{\rm Fr} (q, \omega) = \frac{n e^2}{m} \frac{1}{i \omega + \frac{\eta}{mn (1 + i \tau \omega )}q^2}.
\end{equation}
%
This equation interpolates between the classical fluid in Eq.~\eqref{Eq.transcondCLfull} at $\omega \tau\ll 1$ and the solid in Eq.~(\ref{Eq.transcondCS}) at
$\omega \tau\gg 1$. It contains a modified pole structure that give rise to a propagating shear sound wave with a momentum gap~\cite{Trachenko2015, Lucas2019, Trachenko2020}, 
\begin{equation}\label{Eq.shearFr}
\omega _{\rm Fr} (q) = \frac{i}{2\tau} + \sqrt{c_{\tau}^2 q^2 - \frac{1}{4\tau^2}}, \quad c_{\tau}^2 = \frac{\eta}{nm\tau}.
\end{equation}
Here when $x<0$, we use the convention that $\sqrt{x}=-i\sqrt{|x|}$.
This form is remarkably similar to what we have found for the shear sound in Fermi liquids at finite temperature in Sec.~\ref{Sec.II}.
In fact, in the limit of small momenta, the transverse conductivity of the LFL at finite temperature, which can be obtained by taking $\Gamma_1=0$ from the more general Eq.~\eqref{Eq.condbk}, which we will discuss in the next section, reads as
%
\begin{equation}
\sigma _\perp ^{\rm LFL} (v_{\rm F} q \ll {\rm max}\{\Gamma_2, \omega \},\omega,T) \approx \frac{n e^2}{m} \frac{1}{i \omega + \frac{F_1 +1}{4(\Gamma _2 + i \omega)} v_{\rm F}^2 q^2}.
\label{LFL_visco}
\end{equation}
%
On comparison with Eq.~\eqref{Eq.transcondFr}, one concludes that the time scale in Frenkel’s theory, $\tau$, is simply given by the inverse quasiparticle collision rate $\tau = \Gamma _2 ^{-1}$. 
We emphasize again that the analogy between viscoelastic fluids and LFLs at nonzero $T$ only holds for $F_1\gg 1$, when the pole lies in the regime of validity of Eq.~(\ref{LFL_visco}) far away from the particle-hole continuum. The discrepancy with the classical model is particularly evident at $F=1$, where the spectrum in complex frequency space undergoes a sharp transition to one without propagating collective mode as illustrated in Fig.~\ref{Fig.schematic}.




At low frequencies, $\omega\ll \Gamma_2$, the LFL exhibits a shear diffusion pole with diffusion constant
$D = (1 + F_1) v_{\rm F}^2/2 \Gamma _2 = \eta / m n$ regardless of the value of $F_1$ (cf. Sec.~\ref{Sec.II}), which recovers the well-known divergence with temperature of the classical viscosity of the Fermi fluid~\cite{LifshitzPitaevskii,Pomeranchuk1950,Abrikosov1959,Brooker1968,Jensen1968}.
Such a divergence of the classical viscosity at low temperatures, which is present even in weakly non-interacting Fermi liquids, is a symptom of the emergence of the non-classical behavior of the fluid that we have previously discussed. The fact that the transverse conductivity is dominated entirely by the diffusion pole at finite temperatures and small $(q,\omega)$ can be understood from Fig.~\ref{Fig.schematic}, which shows that the modes making up the particle-hole continuum are completely displaced in the complex-$\omega$ plane to always have a finite imaginary part in their dispersion, even as $q \rightarrow 0$, whereas the shear diffusion pole asymptotes continuously to $(q,\omega)=(0,0)$ and thus dominates the response in such limit. This is ultimately a consequence of the conservation of momentum (when $\Gamma _1 = 0$) which prohibits currents from decaying locally and 
turns them into slow hydrodynamic modes~\cite{ChaikinLubensky}.

\section{Shear resonances in ultraclean channels}\label{Sec.IV}

In this section we develop a theory to describe the dynamics of the LFL in a strip geometry, which will allow us make concrete experimental predictions.
To include boundary effects, we adopt the minimal but realistic model proposed in Ref.~\citenum{LevGregPNAS}, which combines specular boundary conditions with boundary friction modeled as an enhancement of the momentum-relaxing collisions at the boundary of the form
$I [p_{\rm F}] \rightarrow I[p_{\rm F}] + I_{\rm bd}[ p_{\rm F} ]$,
\begin{equation}
I_{\rm bd}[ p_{\rm F}] = b \delta \left(|x| - \frac{W}{2}\right)\left(P_1 [p_{\rm F}] + P_{-1} [ p_{\rm F}] \right),
\end{equation}
where $x \in (-W/2,W/2)$, $y \in (-\infty,\infty)$. As demonstrated in Ref.~\citenum{LevGregPNAS} this model captures the hydrodynamic,  diffusive, and ballistic regimes of metals and their crossovers. For related studies see Refs.~\citenum{LevGregNat, LevGregPNAS, Alekseev2018, Lucas1,Lucas2,Lucas3}. 

We have found an exact analytic solution of the LKE [Eq.~\eqref{Eq.LKE}] for this model with finite Landau parameters $\left\lbrace F_0 ,F_1\right\rbrace$ in addition to all of the above ingredients which we present in the following 
(see Supplemental Material~\cite{supplementary} for details).
Because translation symmetry along $x$ is broken by the presence of the boundaries, the conductivity that determines the current along the channel, $j_y(x,t)$, in response to a driving electric field along the channel, $E_y(x,t)$, is a function of two wave vectors:
\begin{eqnarray}
j_y (q, \omega) &=& \sum_{q'}\sigma _y (q,q',\omega) E _y (q', \omega), \\
\sigma _y (q,q',\omega) &=& \delta _{q, q'}\sigma ^{\rm bk} _y (q,\omega)+ \sigma ^{\rm bd} _y (q,q',\omega).
\end{eqnarray}
The conductivity can be expressed as the sum of a bulk (bk) contribution: 
\begin{align}
&\sigma ^{\rm bk}_y (q, \omega) = \frac{n e^2}{m}\frac{2i z}{ F_1 z ^2 -(v_{\rm F}q)^2  - 2i z \Gamma _2 }, \label{Eq.condbk}\\
&z = \omega-i (\Gamma_1 +\Gamma_2) - \sqrt{\left[\omega-i (\Gamma_1 +\Gamma_2)\right] ^2 -(v_{\rm F}q)^2}, 
\end{align}
and a boundary (bd) contribution:
\begin{eqnarray}
\frac{\sigma ^{\rm bd}_y (q, q',\omega)}{\sigma ^{\rm bk} _y (q,\omega) \sigma ^{\rm bk} _y (q', \omega)} &=& -\frac{ \cos \left(\frac{\pi q}{q_0}\right)\cos \left(\frac{\pi q'}{q_0}\right)  }{\bar{\sigma}^{\rm bd}_y + \bar{\sigma} ^{\rm bk}_{y} (\omega) }, 
\end{eqnarray}
where $q$ is the momentum along  $x$, $q_0=2\pi/W$, $m$ is the transport mass,
$\bar{\sigma} ^{\rm bk}_{y} (\omega) = \sum _{n\in \mathbb{Z}} \sigma ^{\rm bk}_y (n q_0, \omega)$ is the transverse conductivity measuring the bulk response to a periodic array of delta-function perturbations, and $\bar{\sigma} ^{\rm bd} _y = ne^2 W/m b$ parametrizes boundary scattering.
The total conductivity for a uniform driving field is obtained by taking the $q,q' \rightarrow 0$ limit of the above expressions,
\begin{eqnarray}
&&\sigma _y (\omega) = \sigma_{\rm D} (\omega) \left( 1 - \frac{ \sigma _{\rm D} (\omega) }{\bar{\sigma} ^{\rm bd} _y + \bar{\sigma} ^{\rm bk}_{y} (\omega)} \right), \label{Eq.fullcond}
\end{eqnarray}
where $\sigma _{\rm D} (\omega ) = ne^2 /m(i \omega + \Gamma _1)$ is the frequency--dependent Drude conductivity.
The expression in Eq.~\eqref{Eq.fullcond} can be understood as the self-consistent response of the LFL to both an externally applied electric force and the boundary friction.
In a single equation, our solution encompasses the effects of disorder, interactions, as well as boundary scattering, controlled respectively by the parameters $\Gamma_{1,2}$, $F_1$, and $b/W$,  and therefore captures the hydrodynamic, diffusive, ballistic, and LFL regimes on equal footing. Notice that $F_0$ is absent in our expressions because of the absence of density fluctuations for driving electric fields parallel to the channel.

The conductivity in Eq.~\eqref{Eq.fullcond} is shown for a metal with ($F_1 = 3.0$) and without ($F_1 = 0.5$) shear sound in Fig.~\ref{Fig.mainfig}(d). In the former case, there are sharp dips at the shear sound energy, $\omega={\rm Re}\,\omega_s$, evaluated at integer multiples of $q_0$.
In Figs.~\ref{Fig.mainfig}(b) and \ref{Fig.mainfig}(c), we see that the resonant current becomes purely imaginary, i.e., it is out of phase with the applied field. 
Therefore, in the limit of boundary-dominated scattering, metals with shear sound display a \textit{dissipationless} response at the resonant frequencies of this mode. As we will see, this is analogous to the response of a sliding crystal which is subject to friction only at the boundaries.

These conductivity minima acquire finite values in the presence of weak bulk scattering. The electron-electron collision rate is expected to scale as $\Gamma_2=(E_F/2 \pi) \left(k_B T/E_{\rm F}\right)^2$ up to logarithmic corrections~\cite{Hodges1971,Chaplik1971,Bloom1975,eecolrate,Fukuyama1983,Zheng1996,Jungwirth1996,
Menashe1996,Reizer1997,Narozhny2002,novikov2006viscosity} and, therefore, can be easily suppressed by cooling the metal well below the Fermi temperature. The electron-impurity collision rate is limited at low temperatures by the bulk elastic mean-free path, $\lambda =  v_{\rm F}/\Gamma _1$.
We estimate that the shear sound dips would be visible in metals with $\lambda \gtrsim 5W$ at low temperatures. Furthermore, samples with enhanced boundary scattering relative to bulk scattering should lead to more pronounced conductivity dips.

\section{Comparison with an Ideal Crystal Sliding in a Channel}\label{Sec.V}

In this section we would like to illustrate the behavior of a crystal driven by an external uniform force through a clean channel in the presence of enhanced friction at the boundaries.
We demonstrate that the aforementioned dissipationless resonant driving of the Fermi liquid at the harmonics of the shear sound is indeed a hallmark behavior of sliding crystals in such channels. In particular, we will see that in the case of a clean channel with friction arising only from the boundary, the crystal driven at the exact resonant frequency corresponding to the harmonics of its transverse sound self-consistently pins itself with zero velocity at the boundary so as to minimize energy dissipation.

\begin{figure}[t]
\includegraphics[scale=1.0]{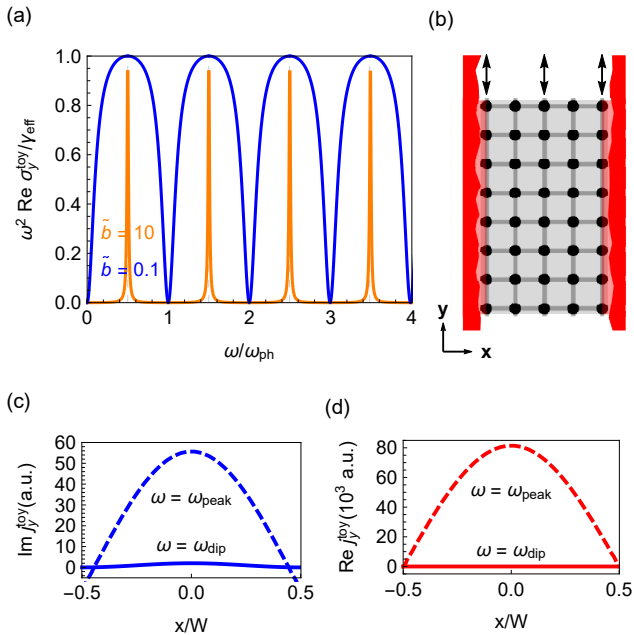}
\caption{\label{Fig.diptopeak} 
(a) Real part of the channel conductivity of the 2D sliding crystal, in units of $\gamma _{\rm eff} / \omega ^2$, for $\tilde{b} = 0.1$ (blue) and $\tilde{b}=10$ (orange), where $\tilde{b}$ is an energy scale parametrizing boundary friction and $\gamma _{\rm eff}$ is an effective scattering rate analogous to $b/W$ and $\Gamma _{\rm eff}$ respectively in the LFL case. All energies are measured in units of the transverse phonon frequency $\omega _{\rm ph}$ (see Supplemental Material~\cite{supplementary} for full model). (b) Schematic of the 2D sliding crystal toy model comprising a tetragonal crystal confined in a channel with only boundary friction (red). (c) Out-of-phase (imaginary) and (d) in-phase (real) current profiles in the crystal in the clean limit $\Gamma _{1,2} \rightarrow 0$. Solid curves correspond to the frequency of the first conductivity dip in (a) ($\tilde{b} = 0.1$) while dashed curves correspond to the frequency at the first conductivity peak in (a) ($\tilde{b} = 10$).}
\end{figure}

To illustrate this, we consider a toy model of a two-dimensional tetragonal crystal~\cite{Kittel} confined in a channel with boundary friction aligned with one of its crystal axes [see Fig.~\ref{Fig.diptopeak}(b)]. The crystal slides in response to an alternating external force along the channel, experiencing friction at the edges analogous to the boundary scattering in the LFL.
Because the translational invariance of the crystal along the infinite direction of the channel ($y$-axis) is preserved during the oscillatory driving, without loss of generality, it is sufficient to consider the motion of a single chain describing a row of $N$ atoms across the channel.
The displacement of each atom from its equilibrium position along $y$ is described by the following equation of motion:
\begin{equation}
\ddot{y}_j = F_j - \kappa (2y_j - y_{j+1} - y_{j-1}) - (\gamma + \gamma _b \delta_{j,-\frac{N}{2}}) \dot{y}_j,
\end{equation}
where $j = -N/2, ... N/2$ labels the $x$-coordinate of the atom, $\kappa$ the shear restoring force constant, $F_j$ the external driving force, $\gamma$ the homogeneous bulk friction, and $\gamma _b$ the boundary friction. The masses of the atoms are set to unity. For simplicity, we have considered the case of periodic boundary conditions along $x$ to highlight the qualitative aspects of the system which are identical to the case with open boundary conditions. Details of the solution of the equations of motion are presented in Section III of the Supplemental Material~\cite{supplementary}, and here we will summarize the resulting behavior.
Figure~\ref{Fig.diptopeak}(a) shows the conductivity, i.e., the average velocity of atoms divided by the external force, of such a sliding crystal. 
In the absence of bulk friction, the real part of the conductivity exhibits zeros at frequencies corresponding to the harmonics of the transverse phonon of the crystal at wavelength $W$.
Resonantly driving the system at these frequencies creates a current profile that is out of phase with the drive: the crystal pins at the boundary and self-consistently avoids energy dissipation in an analogous fashion to the Fermi liquid with shear sound [see Figs.~\ref{Fig.diptopeak}(c) and \ref{Fig.diptopeak}(d)].

When probed optically, the sliding crystal therefore does not exhibit the resonant absorption typical of a crystal with pinned boundaries. The latter scenario can be described as a limiting case of the sliding crystal at infinite boundary friction. Indeed, when the boundary dissipation increases, the dips broaden, ultimately giving rise to resonant peaks at half-integer multiples of the fundamental frequency once the dissipative boundary force exceeds the shear restoring force of the crystal~\cite{supplementary}. 
Such peaks do not have a counterpart in the case of the LFL, where off-resonant pinning at the boundary is prevented by scattering to the incoherent particle-hole continuum. Consequently, the conductivity dips signaling the shear sound in the LFL remain narrow even in the limit of arbitrarily strong boundary scattering~\cite{supplementary}.

\section{Summary and discussion}\label{Sec.VI}

As we have shown, moderately interacting metals display a sharp shear sound collective mode which exists even in the presence of weak impurity and electron-electron collisions. This mode leaves clear fingerprints in clean metallic channels at low temperatures in the form of sharp resonant dips in the conductivity at frequencies controlled by the shear sound dispersion in Eq.~\eqref{Eq.shearcomplexdispphys}, and that resemble the transverse sound resonance of a sliding crystal, despite the metal lacking any form of long-range crystalline order. There already exist various ultra-clean materials that feature a strongly interacting metallic state before a metal insulator transition which are therefore ideal platforms to discover the shear sound.
These include MgZnO/ZnO, Si MOSFETs, AlAs, and p-GaAs~\cite{RevModPhys.82.1743, Dolgopolov2019, Kravchenko2003, Falson2018}. They have been shown to have large mass enhancements and therefore Landau parameters with $F_1>1$~\cite{Solovyev2017, Shashkin2003, Shashkin2002, Shashkin2007, Vakili2004, Falson2018}.
For example, in MgZnO/ZnO two-dimensional electron gases (2DEGs) we estimate that channels of about 1 $\mu$m, at temperatures below 2 K, and with densities so that the quasiparticle mass is enhanced to be larger than twice the bare mass, would display visible shear sound resonances in their conductance at frequencies of about $\omega \sim 0.1$ THz.

\section{Acknowledgements:}

We are thankful to P. S. Alekseev, J. Falson, S. Simon, and A. Chubukov for valuable discussions. We are grateful to an anonymous referee for emphasizing the similarities between shear modes of Fermi liquids at finite temperatures discussed in this work and those derived from the Frenkel model and bringing to our attention the relevant references. P.-Y.C. was supported by the Young Scholar Fellowship Program by Ministry of Science and Technology (MOST) in Taiwan, under MOST Grant for the Einstein Program MOST Grant No. 108-2636-M-007-004.


%

\onecolumngrid

\clearpage
\appendix
\section*{Supplementary Material}

\section{Solving the linearized kinetic equation with bulk and boundary relaxation}\label{Sec.fullsolve}

Using an ansatz of the form $p(x,\theta, t) = p(q,\theta) e^{i(\omega t - q x)}$, the linearized kinetic equation (LKE, main text Eq.~(1)) becomes:
\begin{eqnarray}
\left( i \omega - i v_{\rm F} q \cos \theta + \Gamma _1 + \Gamma _2 \right) p(q,\theta) 
&=& e E_q \sin \theta + \left(i F_0 v_{\rm F} q \cos \theta + \Gamma _1 + \Gamma _2 \right) P_0 (q) \nonumber \\
&& + \left(i F_1 v_{\rm F} q \cos \theta + \Gamma _2 \right) p_1 (q, \theta)  - \sum_{q'} \frac{b q_0 }{2\pi} \cos \left(\frac{\pi (q - q')}{q_0}\right)p_1 (q', \theta), \label{Eq.LLKE}  \nonumber \\
\end{eqnarray}
%
%
where $p_{l>0} (q, \theta) = P_l [p] + P_{-l} [p]$  and $P_0 (q) =P_0 [p]$ with $P_l [p] = e^{i l \theta} \int  \frac{d\theta}{2\pi} e^{-i l \theta} p(q,\theta)$. The Fermi radius then satisfies the expression:
\begin{eqnarray}\label{Eq.collsoln}
&&p(q,\theta) = \frac{-i \tilde{E}_q \sin \theta + \left(F_0 \cos \theta - i \tilde{\Gamma} _1- i \tilde{\Gamma} _2 \right) P_0 (q) + \left(F_1 \cos \theta - i \tilde{\Gamma} _2 \right) p_1 (q, \theta) + i \sum_{q'} \tilde{B}(q, q') p_1 (q', \theta) }{\left( s - \cos \theta - i \tilde{\Gamma} _1 - i \tilde{\Gamma} _2 \right) }, \nonumber \\ \\
&& s = \frac{\omega}{v_{\rm F} q}, \quad \tilde{\Gamma} _i = \frac{\Gamma _i}{v_{\rm F} q}, \quad \tilde{E} _q = e \frac{E _q}{v_{\rm F} q}, \quad \tilde{B}(q, q') = \frac{b q_0}{2\pi v_{\rm F} q} \cos \left(\frac{\pi (q - q')}{q_0}\right),
\end{eqnarray}
To exploit the symmetry of the solutions, we express the Fermi radius explicitly in terms of even and odd Chebyshev polynomials,
%
\begin{eqnarray}
p (q,\theta) &=& \sum_{l=-\infty}^{\infty} P_l (q) e^{i l \theta} = P_0 (q) +  \sum_{l=1}^{\infty} p_l (q,\theta), \\
p _l (q,\theta) &=&  2\left(P_l ^+ (q) \cos \left(l \theta \right) + P_l ^- (q) \sin \left( l \theta \right) \right),  \\
P_l ^{+} (q) &=& \int \frac{d\theta}{2\pi} \cos (l \theta) p  (q,\theta), \quad P_l ^{-} (q) = \int \frac{d\theta}{2\pi} \sin(l \theta) p (q,\theta).\label{Eq.Chebydef}
\end{eqnarray}
The general solution is then solved component by component by projecting Eq.~\eqref{Eq.collsoln} onto the various Cheybyshev sectors using Eq~\eqref{Eq.Chebydef}. This gives rise to a set of coupled self-consistency equations between $P_0 (q)$ and $P_1 ^{\pm} (q)$:

\begin{eqnarray}
P_0 (q) &=& q\left(F_0 \Omega _1  (q) - i \left(\tilde{\Gamma} _1+ \tilde{\Gamma} _2\right)\Omega _0  (q) \right) P_0 (q) +  2q \left(F_1 \Omega _2  (q)- i \tilde{\Gamma} _2 \Omega _1  (q) \right)  P_1^+ (q)  \nonumber \\
&&\quad + i 2q \Omega _1  (q)\sum_{q'} \tilde{B}(q, q')  P_1^+ (q'), \nonumber \\ \\
P_1^+ (q) &=& q\left(F_0 \Omega _2  (q) - i \left(\tilde{\Gamma} _1+ \tilde{\Gamma} _2 \right) \Omega _1  (q) \right) P_0 (q) +  2q \left(F_1 \Omega _3  (q) - i \tilde{\Gamma} _2 \Omega _2  (q) \right)  P_1^+ (q) \nonumber \\
&&\quad  + i 2q\Omega _2  (q)  \sum_{q'} \tilde{B}(q, q')  P_1^+ (q'), \nonumber \\ \\
P_1^- (q) &=& q\Big(\Omega _0  (q)  - \Omega _2  (q) \Big) \left( -i \tilde{E}_q -2i \tilde{\Gamma} _2 P_1^- (q) + i 2  \sum_{q'} \tilde{B}(q, q') P_1^- (q')\right) \nonumber \\
&&\quad +  2  q\Big( \Omega _1  (q)  - \Omega _3 (q)\Big) F_1 P_1^- (q),  \label{Eq.p1m}  \\
\Omega _l  (q)&=& \frac{1}{q}\int \frac{d\theta}{2\pi} \frac{(\cos \theta )^l}{\left(s - \cos \theta - i \tilde{\Gamma} _1 - i \tilde{\Gamma} _2 \right)} = v_{\rm F} \int \frac{d\theta}{2\pi} \frac{(\cos \theta )^l}{\left(\omega  - v_{\rm F} q \cos \theta - i \Gamma _1 - i \Gamma _2 \right)}.
\end{eqnarray}
%
%
%
The solutions of which can then be used to obtain the other $P_{l \geq 2} ^{\pm}(q)$ components. Projecting to the Chebyshev sectors reveals the decoupling between the even ($+$) and odd ($-$) sectors. 

In this work, we are particularly interested in probing the shear collective mode and therefore restrict the following discussion to the solutions of the odd sector, $P_0 (q) = P_l ^+ (q) = 0$. Rearranging Eq.~\eqref{Eq.p1m} leads us to the self-consistent solution
%
\begin{eqnarray}
P_1^- (q) &=&  \frac{1}{2}\tau (q) \left(e E_q- \frac{b q_0}{\pi}\sum _{q'} \cos \left(\frac{\pi (q-q')}{q_0}\right) P_1 ^- (q') \right),  \label{Eq.p1mq} \\
\tau (q) &=&  - i \frac{1}{v_{\rm F} q} \frac{2\Omega _{02}  (q) }{1+2i \Omega _{02}  (q)  \tilde{\Gamma} _2  -  2  \Omega _{13}  (q) F_1}, \\
\Omega _{ij}  (q) &=& q\left(\Omega _{i}  (q)  - \Omega _{j}  (q) \right).
\end{eqnarray}

%
To proceed, we expand the cosine above and introduce the quantities $\mathcal{C}_1^- =  \sum _{q'} \cos \left(\frac{\pi q'}{q_0}\right) P_1 ^- (q')$ and $\mathcal{S}_1^-= \sum _{q'} \sin \left(\frac{\pi q'}{q_0}\right) P_1 ^- (q')$.
Since $q'$ in Eq.~\eqref{Eq.p1mq} are integer multiples of $q_0 = \frac{2\pi}{W}$, the discretization wave-vector set by the channel width $W$, we have $\mathcal{S}^-_1 = \sin \left(\frac{\pi q}{q_0}\right) =0$. Performing $ \sum _{q} \cos \left(\frac{\pi q}{q_0}\right)$ on both sides of Eq.~\eqref{Eq.p1mq} leads to the solution:
%
\begin{eqnarray}
\mathcal{C}_1^-  &=& \frac{\frac{1}{2} \sum _q \cos \left(\frac{\pi q}{q_0}\right)  \tau (q) e E_q}{1+\frac{bq_0}{2\pi } \sum _{q''} \tau (q'')}.
\end{eqnarray}
Substituting this expression back into Eq.~\eqref{Eq.p1mq} gives us the self-consistent solution:
\begin{eqnarray}
P_1^- (q) &=&  \frac{1}{2}e\tau (q) \left( E_q- \frac{b q_0}{2\pi}\cos \left(\frac{\pi q}{q_0}\right) \frac{\sum _{q'} \cos \left(\frac{\pi q'}{q_0}\right)  \tau (q') E_{q'}}{1+\frac{bq_0}{2\pi } \sum _{q''} \tau (q'')}\right).
\end{eqnarray}

Restoring the explicit $\omega$-dependence of the various quantities, the transverse current density is given by:
\begin{eqnarray}\label{Eq.jperpx}
j_y (q,\omega) &=& e\frac{p_{\rm F}^2}{m}\int \frac{d\theta}{(2\pi)^2} \sin (\theta) p (q,\theta) = e\frac{p_{\rm F}^2}{2\pi m} P^-_1 (q,\omega) , \\
&=&\frac{n e^2}{m}\tau (q,\omega) \left( E_y(q,\omega)- \frac{b q_0}{2\pi}\cos \left(\frac{\pi q}{q_0}\right) \frac{\sum _{q'} \cos \left(\frac{\pi q'}{q_0}\right)  \tau (q', \omega) E_y(q',\omega)}{1+\frac{bq_0}{2\pi } \sum _{q''} \tau (q'', \omega)}\right) \nonumber  \\
&=& \sum_{q'}\sigma _y (q,q',\omega) E _y (q', \omega), \\
\sigma _y (q,q',\omega) &=& \delta _{q, q'}\sigma ^{\rm bk} _y (q,\omega)+ \sigma ^{\rm bd} _y (q,q',\omega),\\
\sigma ^{\rm bk}_y (q,\omega) &=& \frac{n e^2}{m} \tau (q,\omega),  \\
\sigma ^{\rm bd}_y (q, q', \omega) &=& -\frac{\cos \left(\frac{\pi q}{q_0}\right)\cos \left(\frac{\pi q'}{q_0}\right) \sigma ^{\rm bk} _y (q, \omega) \sigma ^{\rm bk} _y (q', \omega) }{\bar{\sigma}^{\rm bd}_y + \bar{\sigma} ^{\rm bk}_{y} (\omega)},\\
\bar{\sigma} ^{\rm bk}_{y} (\omega)  &=&  \sum _{n\in \mathbb{Z}} \sigma ^{\rm bk}_y (n q_0, \omega), \quad \bar{\sigma} ^{\rm bd} _y = \frac{ne^2}{m}\frac{W}{b},
\end{eqnarray}
as per Eqs.(9)--(14) in the main text, where we explicitly separate into the bulk and boundary contributions. In going to the second line, we used $n = \frac{p_{\rm F}^2}{4\pi}$, the electron density for a circular Fermi surface. 

For the specific case of a uniform electric field, $E_y( q, \omega) = E_y (\omega) \delta _{q,0}$, the average transverse conductivity is
\begin{eqnarray}
\sigma _y (\omega) = \sigma ^{\rm bk} _y (0,\omega) + \sigma ^{\rm bd} _y (0,0,\omega) =  \sigma_{\rm D} (\omega) \left( 1 - \frac{ \sigma _{\rm D} (\omega) }{\bar{\sigma} ^{\rm bd} _y + \bar{\sigma} ^{\rm bk}_{y} (\omega)} \right), \label{Eq.suppfullcond}
\end{eqnarray}
i.e. Eq.(14) in the main text. Here we derive the frequency-dependent Drude conductivity explicitly:
\begin{eqnarray}
\sigma _D (\omega) &=& \sigma ^{\rm bk}_y (0,\omega) = \frac{n e^2}{m} \tau (0,\omega),  \\
&=& -\frac{n e^2}{m} \frac{2 i\left(\Omega _{0}  (0,\omega) - \Omega _{2}  (0,\omega) \right)}{v_{\rm F}+2 i\left(\Omega _{0}  (0,\omega) - \Omega _{2}  (0,\omega) \right)\Gamma _2}, \quad \Omega _l  (0, \omega) = \frac{v_{\rm F}}{\omega - i(\Gamma _1 + \Gamma _2)}\int \frac{d \theta}{2\pi} (\cos \theta)^l ,  \nonumber \\
&=& \frac{ne^2}{m(i \omega + \Gamma _1)}. \nonumber
\end{eqnarray}

Finally, the spatial current profiles can be obtained via the inverse Fourier transform: 
\begin{eqnarray}
j_y (x, \omega) &=& \sum _q e^{- i q x} j_y (q, \omega) 
= \sum _q e^{- i q x}  \sum_{q'}\sigma _y (q,q',\omega)  E_y (\omega)  \delta _{q',0}, \nonumber \\
&=& E_y (\omega)  \sigma _D (\omega ) \left(1 -  \frac{\sum _{l\in \mathbb{Z}} e^{- i l q_0 x} (-1)^l \sigma ^{\rm bk}_y  (l q_0,\omega)}{\bar{\sigma} ^{\rm bd} _y + \bar{\sigma} ^{\rm bk}_{y} (\omega)} \right).
\end{eqnarray}

\clearpage

\section{Shear Sound Complex Dispersion in the Infinite System}\label{Sec.complexshear}

In this section we outline the derivation of the generic complex dispersion relation for the shear sound in the infinite system ($b = 0$) in the absence of external fields, but with finite Landau parameters and scattering. The dispersive and purely decaying solutions, as well as the Pomeranchuk instability discussed in the main text can be obtained from the different regions in parameter space for which the solution is valid.

\begin{figure}[t]
\includegraphics[scale=1.0]{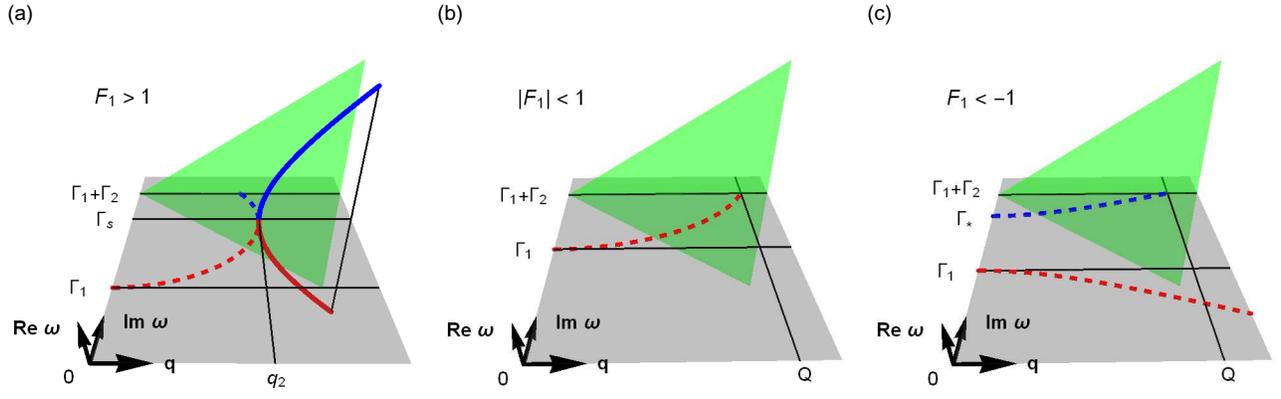}
\caption{\label{Fig.complexsheardisp}Schematic showing qualitatively distinct solutions of the shear collective mode dispersion $\omega _+$ (blue) and $\omega _-$ (red) for the three regimes of $F_1$ values indicated. Dispersive solutions only exists for $F_1 > 1$ shown in (a) and are plotted in solid lines while decaying modes are plotted in dashed lines. (c) Pomeranchuk instability: $\omega _-$ describes an exponentially growing solution. Particle-hole continuum is given by the green region. $\Gamma _s = \Gamma _1 + s_1 \frac{\Gamma _2}{\sqrt{F_1}}$ and $\Gamma _{p-h} = \Gamma _1 + \Gamma _2$.}
\end{figure}

Primarily, we solve Eq.~\eqref{Eq.p1m} with $b = 0$, or equivalently, for the zeros of $\tau ^{-1} (q, \omega)$,
\begin{eqnarray}\label{Eq.complexshearcond}
1+2i \Omega _{02} (q, \omega)\tilde{\Gamma} _2 - 2 F_1 \Omega _{13}  (q, \omega) = 0.
\end{eqnarray}
The integrals $\Omega _{ij}  (q, \omega) = \Omega _{ij}  (\zeta)$ are functions of the single variable $\zeta = s - i( \tilde{\Gamma}_1 + \tilde{\Gamma}_2)$, where we restrict to the case $ \tilde{\Gamma}_{1,2} > 0$. We extend $s$ into the complex plane, $s \rightarrow \mathfrak{s} = s + i \gamma$, and evaluate these integrals by a change of variables $z = e^{i \theta}$, 
%
\begin{eqnarray}
\Omega _{02} (\zeta) &=& \int \frac{d\theta}{2\pi} \frac{\sin ^2 \theta }{\zeta - \cos \theta } =-i \frac{1}{4\pi} \oint _{C} dz \frac{(z^2 - 1)^2}{z^2 (z-z_+)(z-z_-)} 
= \left\lbrace \begin{array}{c}
z_-, \quad s > 0, \\
z_+, \quad s < 0, 
\end{array}\right. \\
\Omega _{13} (\zeta) &=& \int \frac{d\theta}{2\pi} \frac{\cos \theta \sin ^2 \theta }{\zeta - \cos \theta } =-i \frac{1}{8\pi} \oint _{C} dz \frac{(z^2 - 1)^2(z^2 + 1)}{z^3 (z-z_+)(z-z_-)} 
= \left\lbrace \begin{array}{c}
\frac{1}{2} z^2_-, \quad s > 0, \\
\frac{1}{2} z^2_+, \quad s < 0, 
\end{array}\right. \\
z_{\pm} &=& \zeta \pm \sqrt{\zeta ^2-1}, \quad z_+z_- = 1,
\end{eqnarray}
%
where $C$ denotes the unit circle. The above results are valid for $|z_+| \neq |z_-| \neq 1$, corresponding to solutions outside the particle-hole continuum. Substituting them into Eq.~\eqref{Eq.complexshearcond} and self-consistently solving for $\omega$, we arrive at the following solutions 
\begin{equation}\label{Eq.shearcomplexfull}
\omega _\pm = i \left( \Gamma _1 + s_1 \frac{\Gamma _2}{\sqrt{F_1}}\right) \pm s_1 \sqrt{(v_{\rm F} q)^2 - \frac{\Gamma _2^2}{F_1}}, \quad s_1 = \frac{1+F_1}{2\sqrt{F_1}}
\end{equation}
which are only valid for the following regions in parameter space with $q \geq 0$:
\begin{itemize}
\item
$\omega  _+$ solution only exists for $|F_1| > 1$ and for the following wave-vectors:
\begin{align}
& 0 \leq v_{\rm F} q < -\frac{2 \Gamma _2}{1+F_1}, \quad F_1 < -1, \\
& \frac{2 \Gamma _2}{1+F_1} \leq v_{\rm F} q, \hspace{3.7em} F_1 > 1,
\end{align}
\item
$\omega_-$ solution exists for all $F_1$ and for the following wave-vectors:
\begin{align}
& 0 \leq v_{\rm F} q, \hspace{5.4em} |F_1| > 1, \\
& 0 \leq v_{\rm F} q \leq \frac{2\Gamma _2}{1+F_1}, \quad |F_1| \leq 1,
\end{align}
\end{itemize}
These results are summarized in Fig.~\ref{Fig.complexsheardisp}, highlighting the three regimes of $F_1$ values with qualitatively distinct solutions.

\clearpage

\section{Resonant conductivity zeros of the sliding crystal}\label{Sec.toy}

In this section we discuss in greater detail the sliding crystal toy model to illustrate the phenomenon of resonant conductivity dips.

Consider the propagation of elastic waves in a two-dimensional crystal tetragonal crystal. For a wave vector pointing along one of the principal axes of the crystal, entire lines of atoms move in phase with displacements either parallel (longitudinal) or perpendicular (transverse) to the direction of the wave vector. For simplicity, we assume that a line $j$ with displacement $u_j$ experiences a restoring force $F^{{\rm res}} _j$ only from its adjacent lines $j \pm 1 $ that is linearly proportional to the difference of their displacements~\cite{Kittel},
\begin{eqnarray}
F^{{\rm res}} _j = - \kappa (u_j - u_{j+1}) - \kappa (u_j - u_{j+1}).
\end{eqnarray}
Here $\kappa$ is the force constant.

Let us now confine the tetragonal crystal in a channel (see Fig.~3(b) of the main text). For an $N$-atom wide channel, the two columns of atoms at the edges $x _{\pm N/2}$ experience boundary friction $\gamma _b$ due to the roughness of the channel. In addition, we introduce a homogeneous bulk friction $\gamma$, and further drive the system uniformly along $\hat{y}$ parallel to the channel with an external force $F_j \propto e^{i \omega t}$. By symmetry, only the transverse waves are excited by the drive so that we write explicitly $u_j = y_j$.

To highlight the qualitative aspects of the system, we outline below the solution for the case of periodic boundary conditions along $x$ so that $x_j = x_{j + N}$. 
Setting the particles' masses to unity, we have the following equation of motion:
\begin{eqnarray}
\ddot{y}_j &=& F_j - \kappa (y_j - y_{j+1}) - \kappa (y_j - y_{j+1}) - (\gamma + \gamma _b \delta_{j,-\frac{N}{2}}) \dot{y}_j. 
\end{eqnarray}
Using the ansatz $y_j \propto e^{i \omega t}$, the equation of motion in Fourier space reads
%
%

\begin{eqnarray}
\mathcal{F}_q &=& \sum_j e^{-i \frac{2\pi q}{N} j} \left\lbrace \frac{1}{N}\sum _k \mathcal{Y}_k e^{i \frac{2\pi k}{N} j} \left( -\omega ^2 + 2\kappa + i \omega \gamma + i\omega \gamma _b \delta_{j,-\frac{N}{2}} \right) - \frac{\kappa}{N}\sum _k \mathcal{Y}_k e^{i \frac{2\pi k}{N} j} \left( e^{i \frac{2\pi k}{N}} + e^{-i \frac{2\pi k}{N}} \right)\right\rbrace \nonumber \\
&=& \mathcal{Y}_q \left\lbrace -\omega ^2 + 2\kappa \left(1-\cos \left(\frac{2\pi q}{N}\right)\right) + i \omega \gamma \right\rbrace + i \omega \gamma _b \frac{1}{N}(-1)^q \sum _k (-1)^k \mathcal{Y}_k,
\end{eqnarray}
where $\mathcal{F}_q = \sum_j e^{-i \frac{2\pi q}{N} j}F_j$ and $\mathcal{Y}_q = \sum_j e^{-i \frac{2\pi q}{N} j}y_j$ are the respective Fourier components of $F_j$ and $y_j$.


We rewrite the above in a form similar to Eq.~\eqref{Eq.p1mq},
\begin{eqnarray}
\mathcal{Y}_q &=& \eta_{\rm toy} (q, \omega ) \left( \mathcal{F}_q - i \omega \gamma _b \frac{1}{N}(-1)^q \sum _k (-1)^k \mathcal{Y}_k \right), \\
\eta_{\rm toy} (q, \omega ) &=& 
\frac{1}{-\omega ^2 + 2\kappa \left(1-\cos \left(\frac{2\pi q}{N}\right)\right) + i \omega \gamma}, 
\end{eqnarray}
and proceed to solve it in a similar fashion outlined earlier to find
%
\begin{eqnarray}
\mathcal{Y}_q &=& \eta_{\rm toy} (q, \omega ) \left( \mathcal{F}_q - i \omega \gamma _b \frac{1}{N}(-1)^q \frac{\sum_k (-1)^k \eta _{\rm toy} (k, \omega ) \mathcal{F}_k}{1 + i\omega \gamma _b \frac{1}{N}\sum_{q'} \eta _{\rm toy} (q', \omega )} \right). 
\end{eqnarray}

%
%

For a spatially uniform external force $F_j = F$, $\mathcal{F}_q = FN \delta _{q,0}$, an analogous average transverse conductivity relating the average transverse velocity density to the external force can be defined,
\begin{eqnarray}\label{Eq.toycond}
\sigma _{\rm toy} (\omega) &=& \frac{\sum_j \dot{y}_j}{NF} = \frac{i \omega \mathcal{Y}_0}{NF}
= \sigma_{\rm toy, D} (\omega) \left( 1 - \frac{\sigma _{\rm toy, D} (\omega) }{\bar{\sigma} ^{\rm bd}_{\rm toy} + \bar{\sigma} ^{\rm bk}_{\rm toy} (\omega)} \right),
\end{eqnarray}
with the quantities for the toy model analogous to the Fermi liquid,
\begin{eqnarray}
\sigma _{\rm toy, D} (\omega ) &=&  i \omega \eta_{\rm toy} (0,\omega) =\frac{1}{i \omega + \gamma }, \\
\bar{\sigma} ^{\rm bk}_{\rm toy} (\omega) &=& \sum _q \sigma _{\rm toy} ^{\rm bk} (q, \omega) = \sum _q i \omega \eta _{\rm toy} (q, \omega), \\
\bar{\sigma} ^{\rm bd} _{\rm toy}(\omega) &=& \frac{N}{\gamma _b}. 
\end{eqnarray}

The collective modes of this toy model analogous to the shear sound is simply the the transverse phonon, whose positive frequency dispersion is given by the poles of $\lim _{\gamma \rightarrow 0} \eta _{\rm toy} (q, \omega )$,
\begin{eqnarray}
\omega _0 (q) &=& \sqrt{2\kappa \left(1-\cos \left(\frac{2\pi q}{N}\right)\right)}, \quad q \in \mathbb{Z},  \\
\lim _{N\rightarrow \infty} \omega _0 (q) &\simeq & \omega _{\rm ph} q, \quad \omega _{\rm ph} = \sqrt{\kappa} \left(\frac{2\pi}{N}\right),
\end{eqnarray}
where $\omega _{\rm ph}$ denotes the fundamental frequency of the transverse phonon. Rewriting the bulk conductivity, we find
\begin{eqnarray}
\sigma _{\rm toy} ^{\rm bk} (q, \omega) &=& i \omega \eta _{\rm toy} (q, \omega) = \frac{-i \omega}{2 \omega _{\gamma}(q)} \left( \frac{1}{\omega - \omega _{\gamma}(q) - i \frac{\gamma}{2}} - \frac{1}{\omega + \omega _{\gamma}(q) - i \frac{\gamma}{2}} \right), \\
\omega _{\gamma} (q) &=& \sqrt{\omega _0 ^2 (q) - \frac{1}{4}\gamma ^2}, 
\end{eqnarray}
whose real component is 
\begin{eqnarray}
\sigma _{\rm toy} ^{\rm bk '} (q, \omega) &=& \frac{\gamma \omega}{4 \omega _{\gamma}(q)} \left( \frac{1}{(\omega - \omega _{\gamma}(q))^2 + \frac{1}{4}\gamma ^2} - \frac{1}{(\omega + \omega _{\gamma}(q))^2 + \frac{1}{4}\gamma ^2} \right).
\end{eqnarray}
This is none other than the relaxation broadened Lorenzian peak centered at the resonant frequencies $\pm \omega _\gamma (q)$ of the (damped) transverse phonon whose peak value diverges as $\frac{1}{\gamma}$.

The average conductivity of the channel itself however does not necessarily follow the behavior of the bulk conductivity depending on the value of the boundary friction $\gamma _b$. To see this, we rewrite the real part of Eq.~\eqref{Eq.toycond} to leading order in $\frac{\gamma}{\omega}$,
\begin{eqnarray}
\sigma ' _{\rm toy} (\omega) &\simeq & \frac{1}{\omega ^2}\left( \gamma + {\rm Re} \left[ \beta _{\rm toy} (\omega) \right] \right) + \mathcal{O} \left( \frac{\gamma}{\omega}\right)^2, \\
\beta _{\rm toy} ^{-1} (\omega) &=& \frac{N}{\gamma _b} + \bar{\sigma} ^{\rm bk} _{\rm toy} (\omega).
\end{eqnarray}
The quantity $\bar{\sigma} ^{\rm bk} _{\rm toy} (\omega)$ is therefore a series of Lorentzian broadened resonant peaks centered about $\omega _{\gamma} (q \in \mathbb{Z})$ (a Dirac comb in the $\gamma \rightarrow 0$ limit). 
Consequently, there are three qualitatively distinct regimes:
\begin{itemize}
\item
No resonance features: $\gamma \gg \frac{\gamma _b}{N}$.

In this case $ \beta _{\rm toy}^{-1}  (\omega)  \simeq \frac{N}{\gamma _b}$ so that in general, $\sigma ' _{\rm toy} (\omega) \simeq  \frac{1}{\omega ^2}\gamma $.

\item
Resonant dips: $\gamma \ll \frac{\gamma _b}{N} \ll \omega _{\rm ph} \Rightarrow \gamma _b \ll 2\pi \sqrt{\kappa} $.

In general, $ \beta _{\rm toy} ^{-1} (\omega)  \simeq \frac{N}{\gamma _b}$ so that $\sigma ' _{\rm toy} (\omega) \simeq  \frac{1}{\omega ^2}\frac{\gamma _b}{N} $. On resonance however, $ \beta _{\rm toy}^{-1}  (\omega)  \simeq \frac{1}{\gamma}$, so that $\sigma ' _{\rm toy} (\omega) \simeq  \frac{2}{\omega ^2}\gamma \rightarrow 0$ as $\gamma \rightarrow 0$, i.e. resonant dips.

\item
Resonant peaks: $\gamma \ll \omega _{\rm ph} \ll \frac{\gamma _b}{N} \Rightarrow \gamma _b \gg 2\pi \sqrt{\kappa} $.

In this case, the contribution from the resonant peaks become comparable and the typical conductivity becomes roughly constant. At a frequency midway between two adjacent resonant frequencies $\omega _0 (q)$  and $\omega _0 (q+1)$ , their imaginary parts cancel. A narrow peak develops, whose height and inverse width is approximately given by $\min \left(\frac{\gamma _b}{N} , \frac{\omega ^2 _{\rm ph}}{\gamma}\right)$.
\end{itemize}

The second regime corresponds to strong pinning of the crystal at the boundary. The average channel conductivity recovers the resonant peak structure of the bulk conductivity but at shifted frequencies,
\begin{equation}
\omega _{\rm peak} (q) = \sqrt{2\kappa \left(1-\cos \left(\frac{2\pi}{2N}(2q -1)\right)\right)}, \quad q \in \mathbb{Z}.
\end{equation}
These are none other than the odd harmonics of the transverse phonon whose fundamental wavelength is now twice the channel width (c.f. particle in a box). Only the odd harmonics are excited by the spatially uniform driving field due to the symmetry about $x=0$. The crossover between the two regimes is illustrated in Fig.~\ref{Fig.suppdiptopeak}(a) in Sec.~\ref{Sec.visibility}.

\clearpage

\section{Visibility of conductivity dips}\label{Sec.visibility}

In this section, we discuss the effect of bulk relaxation $\Gamma _{1,2}$ on the visibility of the conductivity dips and provide an estimate on the sample cleanliness required to observe them. We are therefore primarily interested in the regime when the shear sound exists as a dispersive mode, $F_1 > 1$.


The momentum conserving relaxation rate $\Gamma_2$ has two different effects on the shear sound, as is evident from Eq.~\eqref{Eq.shearcomplexfull}: it contributes to the broadening with a prefactor $\frac{s_1}{\sqrt{F_1}} = \frac{1+F_1}{2F_1}$ which ranges from $\frac{1}{2}$ to 1 in the regime of interest and it introduces a second order correction to the shear sound dispersion. The latter is rather unimportant when relaxation is weak and we ignore it here. Hence, we focus exclusively on $\Gamma=\Gamma_1$, setting $\Gamma_2\to 0$ in the following, with the implicit assumption that $\Gamma_2$ has the same qualitative effect.

Let us first consider the effect of $\Gamma$ on the bulk transverse conductivity $\sigma _y ^{\rm bk} (q, \omega)$. Using the results in Section~\ref{Sec.complexshear}, we find
\begin{eqnarray}
\sigma _y ^{\rm bk} (q, \omega) &\simeq & \frac{ne^2}{m} \frac{2i z}{v_{\rm F}q(F_1 z^2 -1)} , \quad z = s - i \tilde{\Gamma}- \sqrt{\left(s - i \tilde{\Gamma} \right) ^2 - 1} = z' + i z'',
\end{eqnarray}
where $s = \frac{\omega}{v_{\rm F}q}$, $\tilde{\Gamma} = \frac{\Gamma}{v_{\rm F}q}$, and $z'$ and $z''$ denote respectively the real and imaginary parts of $z$. When $\omega < v_{\rm F} q$ the parameter $z$ is complex even with $\Gamma = 0$. Its magnitude is approximately constant over this range of $\omega$. Physically, this corresponds to contributions coming from the incoherent particle-hole excitations. The functional dependence of $z$ on $\omega$ is qualitatively different for $\omega > v_{\rm F}q$, in which case we can write explicitly
\begin{eqnarray}
&& \sigma _y ^{\rm bk}  (q, \omega > v_{\rm F} q) \simeq \frac{ne^2}{m}\frac{2i z}{v_{\rm F}q F_1} \frac{1}{ \left(z'-F_1^{-\frac{1}{2}} +i z''\right)\left(z'+F_1^{-\frac{1}{2}}  + i z''\right)}, \\ 
&& z' = s - \sqrt{s^2 - 1} + \mathcal{O} (\tilde{\Gamma}^2), \quad  z'' 
= \tilde{\Gamma} \left( \frac{z'}{s-z'} \right) + \mathcal{O} (\tilde{\Gamma}^2), 
\end{eqnarray}
Here, we have expanded $z$ to leading order in $\tilde{\Gamma}$. A resonance occurs when $z' = F_1 ^{-\frac{1}{2}}$, corresponding to $\omega = \omega _{\rm shear} =v _{\rm shear}q$ with $v _{\rm shear}= v_{\rm F}(1+F_1)/2\sqrt{F_1}$. The real part of the conductivity has the form of a Lorentzian in the vicinity of the shear sound frequency 
\begin{eqnarray}\label{Eq.shearbulkcond}
{\rm Re}\,\sigma _y ^{\rm bk }  (q, \omega > v_{\rm F} q) &\simeq & \frac{ne^2}{m}\frac{1}{v_{\rm F}q F_1}  \left( \frac{z''}{(z' -\xi)^2 + z''^2} + \frac{z''}{(z' + \xi)^2  + z''^2} \right),
\end{eqnarray}
analogous to the transverse phonon of the toy model considered in Section~\ref{Sec.toy}. 
To leading order in $\Gamma/\omega$, the real component of the average transverse conductivity Eq.~\eqref{Eq.suppfullcond} is then
%
%
\begin{eqnarray}
{\rm Re}\, \sigma  _y (\omega) &\simeq & \frac{n e^2}{m}\frac{1}{\omega ^2}\left( \Gamma + {\rm Re} \left[ \beta (\omega) \right] \right) + \mathcal{O} \left( \frac{\Gamma}{\omega}\right)^2, \\
\beta^{-1} (\omega) &=& \frac{W}{b} + \frac{m}{ne^2}\sum _{l \in \mathbb{Z}} \sigma _{y} ^{\rm bk} (l q_0, \omega).
\end{eqnarray}
The sum over $l$ has contributions both from the shear sound and the particle-hole continuum.
When $\omega$ is near the first shear-sound resonance, $\omega\simeq v_{\rm shear}q_0$, we can approximate $\beta^{-1} $ by
\begin{equation}
\beta^{-1} (\omega) \simeq \frac{W}{b} -i\Bigl(  \frac{F_1 - 1}{2 F_1}\Bigr)\frac{1}{\omega -  v _{\rm shear} q_0 - i \Gamma} + \zeta _{\rm p-h},
\end{equation}
where $\zeta _{\rm p-h} \sim \log n_{\rm max}W/v_{\rm F}$ is a complex number that is approximately constant for small deviations of $\omega$ and $n_{\rm max} = q_{\rm max}/ q_0$ depends on the large momentum cutoff $q_{\rm max}$, which is determined by the characteristic length scale of the boundary friction or by the Fermi momentum $k_{\rm F}$. 
For simplicity we assume $\log n_{\rm max}$ to be of order one and  drop it in the following.

\begin{figure}[h]
\includegraphics[scale=1.0]{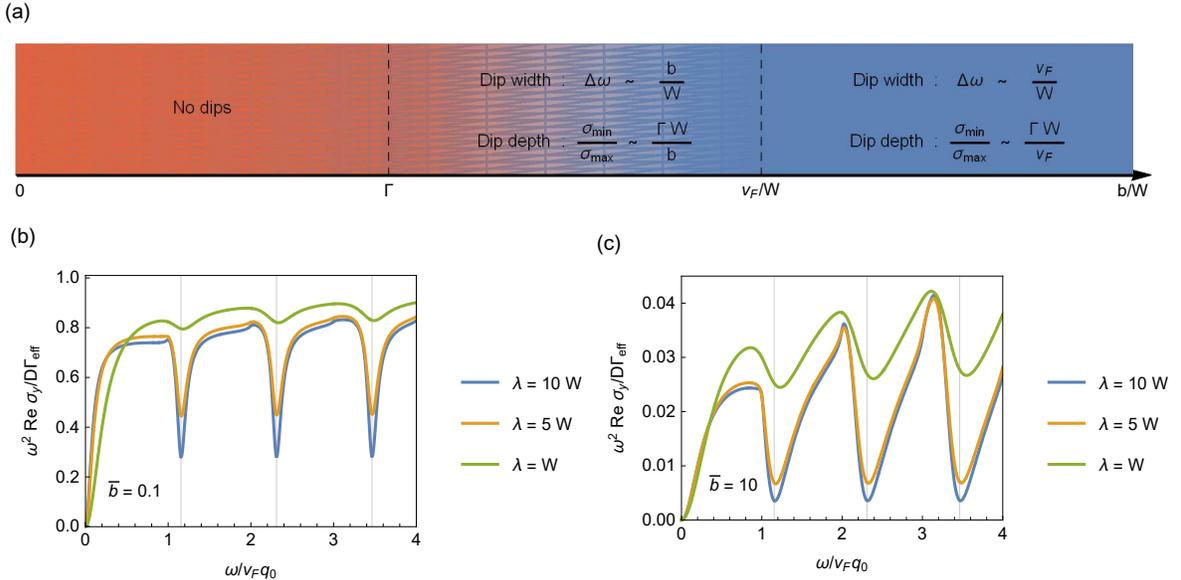}
\caption{\label{Fig.dipwidth} (a) Schematic of the visibility of the conductivity dips as a function of the effective boundary scattering rate $b/W$. The dips appear for $b/W>\Gamma$ and the width and depth saturate for $b/W>v_{\rm F}/W$. (b, c) Plots of the real part of the channel conductivity for different values of the mean free path $\lambda$. The boundary scattering rate is $\bar{b}=0.1$ in (b), and $\bar{b}=10$ in (c) demonstrating the broadening of the dips with stronger boundary scattering.}
\end{figure}

Similar to the case of the toy model, signatures of the shear sound resonance become discernible once the bulk relaxation is smaller than all the other energy scales present, namely $\Gamma \ll b/W,v_{\rm F}/W$. In this regime, the real conductivity becomes 
\begin{eqnarray}\label{Eq.condcontrast}
{\rm Re}\,\sigma  _y (\omega) &\sim & \frac{n e^2}{m}\frac{1}{\omega ^2} \left\lbrace
\begin{array}{cc}
\min(b/W,v_{\rm F}q_0), & \delta \omega \gg \min(b/W,v_{\rm F}/W) \\
\Gamma , & \delta \omega \ll \min(b/W,v_{\rm F}/W)
\end{array}\right. ,
\end{eqnarray}
where $\delta \omega=\omega-\omega_{\rm shear}$. At the resonance, the conductivity has a dip, whose width and depth are controlled by the parameter $\min(b/W,v_{\rm F}/W)$ as illustrated in Fig.~\ref{Fig.dipwidth}(a). We conclude that the dip should be visible in an experiment once the experimental resolution and the energy broadening due to disorder and electron collisions is smaller than both the effective boundary scattering rate $b/W$ and the typical energy of excitations $v_{\rm F}/W$. Similar estimates hold for the resonances at higher harmonics of the shear-sound frequency, $\omega=l\omega_{\rm shear}$. In practice, samples with a mean free path of $\lambda >5W$ should be sufficient to support well developed dips as shown in Fig.~\ref{Fig.dipwidth}(b) and (c).

\begin{figure}[h]
\includegraphics[scale=1.0]{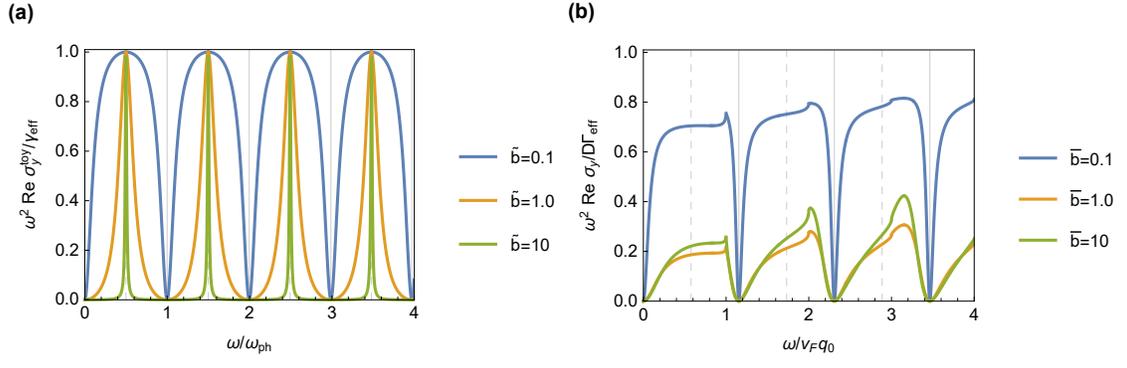}
\caption{\label{Fig.suppdiptopeak} Plots of the real part of the channel conductivity with increasing values of boundary friction showing (a) the broadening of conductivity zeros and the development of resonant peaks in the toy model, and (b) the obstruction of this phenomenon in the LFL due to the presence of the particle-hole continuum. Boundary friction values are $\bar{b}$ (or $\tilde{b}$ for the toy model) = 0.1 (blue), 1 (orange) and 10 (green). Bulk friction is set to $\gamma=\Gamma=0$ in both models. Parameters for the toy model are chosen to emulate the dips at the shear sound resonant frequencies in the LFL with $F_1 = 3$. }
\end{figure}

While the widths of the dips are determined by the boundary scattering rate $b/W$, they saturate to a finite value $\sim v_{\rm F}/W$ in the limit of strong boundary scattering, in which case the widths are typically of the order of the distance between neighboring dips $\sim v_{\rm F}/W$. This is in contrast to the toy model of a sliding crystal, where the width of the dips continue to grow with increasing boundary scattering until they evolve into conductivity peaks located at the mid point between two resonances. This difference is illustrated by a comparison of the Fermi liquid and the toy model in Fig.~\ref{Fig.suppdiptopeak}. In this figure (also in Fig.~\ref{Fig.dipwidth}(b) and Fig.~\ref{Fig.dipwidth}(c) and in main text Fig.~1(d) and Fig.~3(a)), the conductivities are respectively normalized by
\begin{eqnarray}
\sigma _0 (\omega) &=& \frac{ne^2}{m}\frac{1}{\omega ^2}\text{Re}\left[\beta \left(\omega \gg \Gamma, \frac{b}{W} \right)\right] \simeq \frac{D \Gamma _{\rm eff}}{\omega ^2}, \quad D = \frac{ne^2}{m}, \\
\Gamma _{\rm eff} &=& \Gamma + v_{\rm F}q_0 \min \left(\bar{b} , 1\right), \quad \bar{b} = \frac{b}{W v_{\rm F}q_0} = \frac{b}{2\pi v_{\rm F}},
\end{eqnarray}
for the LFL, with $D$ denoting the Drude weight, and
\begin{eqnarray}
\sigma ^{\rm toy} _0 (\omega ) &=& \frac{1}{\omega ^2} \text{Re}\left[\beta _{\rm toy} (\omega \gg \gamma, \gamma _b)\right] = \frac{\gamma _{\rm eff}}{\omega ^2}, \\
\gamma _{\rm eff} &=& \gamma + \omega _{\rm ph} \tilde{b}, \quad \tilde{b} = \frac{\gamma _b}{N\omega _{\rm ph}},
\end{eqnarray}
for the toy model. The dimensionless friction parameter $\bar{b}$ for the LFL is defined relative to $v_{\rm F}q_0$, while its analogue for the toy model $\tilde{b}$ is defined relative to the fundamental transverse phonon frequency.
The parameters $\Gamma _{\rm eff}$ and $\gamma _{\rm eff}$ characterize respectively the effective scattering rates in the LFL and the toy model.
The saturation of the dip width in the case of the Fermi liquid arises from the additional contribution to the conductivity due to particle-hole excitations. Unlike the toy model, it obstructs the complete formation of the resonant peaks.
Indeed, in the somewhat artificial limit of $F_1\to \infty$, the contribution of the particle-hole continuum is diminished and the Fermi liquid becomes more similar to the toy model.

\end{document}